\documentclass[usenatbib]{iau} 
\usepackage{graphicx}

\usepackage{natbib}

\title[Masers in NGC\,253] 
{The Maser-Starburst connection in NGC\,253}

\author[Simon Ellingsen]   
{Simon P. Ellingsen$^1$}

\affiliation{$^1$School of Physical Sciences, University of Tasmania, Hobart 7001, TAS, Australia}

\pubyear{2017}
\volume{336}  
\jname{Astrophysical Masers: Unlocking the Mysteries of the Universe}
\editors{A. Tarchi, M.J. Reid \& P. Castangia, eds.}
\begin{document}

\maketitle

\begin{abstract}
NGC\,253 is one of the closest starburst galaxies to the Milky Way and as such it has been studied in detail across the electromagnetic
spectrum.  Recent observations have detected the first extragalactic class~I methanol masers at 36 and 44 GHz and the first extragalactic HC$_3$N (cyanoacetylene) masers in this source.  Here we discuss the location of the masers with respect to key morphological features within NGC\,253 and the association between the masers and the ongoing starburst.

\end{abstract}

\firstsection 
\section{Introduction}
Masers have proven to be a powerful tool for studying the kinematics of both galactic and extragalactic
environments \citep[e.g.][]{Goddi+11,Miyoshi+95}.  Megamasers are extragalactic maser sources which
have an isotropic luminosity a milliion times or greater than that of typical galactic maser emission from the
same transitions.  Megamaser emission from the 22 GHz transition of water and the 1667 MHz transition of
OH have been known for more than 30 years \citep{DosSantos+79,Baan+82} and there are now more than 100 sources known 
\citep[see the review talk by][in these proceedings]{Henkel17}.  The discovery of strong and common galactic methanol maser transitions
in the 1980s and 1990s lead to searches for methanol megamasers, particularly from the 6.7 GHz transition, however, these
were all unsuccessful \citep{Ellingsen+94,Phillips+98a,Darling+03}.  

The strongest Galactic class~I methanol maser transitions are the $7_{0} \rightarrow 6_{1}A^{+}$ and $4_{-1} \rightarrow 3_{0}E$ transitions (rest 
frequencies of 44 and 36 GHz respectively) and in the first international conference dedicated to astrophysical masers, held in 1992 in Arlington, Andrej Sobolev suggested that the 36~GHz transition may provide the best prospects for being detected in extragalactic sources.  
Furthermore, his subsequent paper \citep{Sobolev93} made the prediction that emission with a strength of around 50~mJy would be present in the nearby starburst 
galaxy NGC\,253.  At the time that Sobolev's paper was written there were relatively few radio telescopes capable of making observations at 
36~GHz and so the prediction remained untested until 2014, when observations by \citet{Ellingsen+14} detected 36~GHz methanol emission with 
a peak intensity of around 20~mJy towards NGC\,253 (Fig.~\ref{fig:ngc253}).  The 36~GHz methanol emission in NGC\,253 is 4 to 5 orders of 
magnitude stronger than that observed in typical Galactic star formation regions and offset from centre of the galaxy by 200--300~pc, along the 
direction of the galactic bar \citep{Ellingsen+14}.

\section{NGC\,253}
\begin{figure}[h]
\begin{center}
 \includegraphics[width=3.4in]{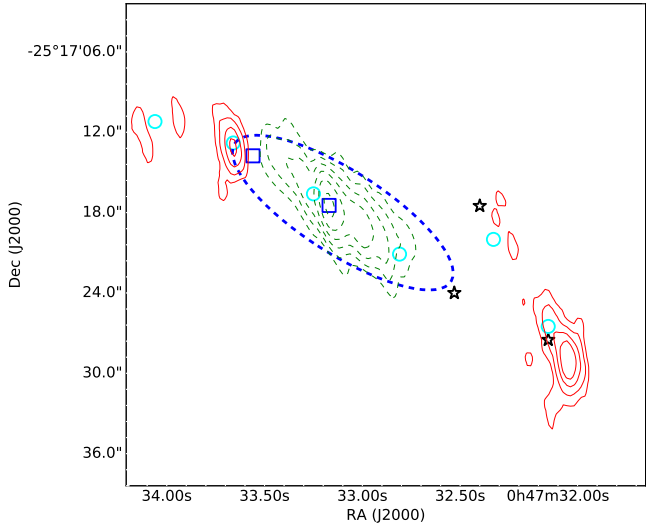} 
 \caption{Integrated 36 GHz methanol maser emission (solid contours) and 7mm radio continuum (dashed contours) towards NGC\,253.  The thick-dashed ellipse marks the half-maximum intensity of the CMZ \citep{Sakamoto+11}.  The squares mark the location of 22 GHz water masers \citep{Henkel+04}, the stars are supershells identified in the CO observations of \citet{Sakamoto+06}, the circles are NH$_3$ cores from \citet{Lebron+11}.}
   \label{fig:ngc253}
\end{center}
\end{figure}

NGC\,253 is the largest galaxy in the Sculptor group, the closest group of galaxies beyond the local group.  It is a barred spiral which we observe at a high inclination angle to our line-of-sight and has a star formation rate several times higher than that of the Milky Way \citep[e.g.][]{Ott+05}, more than half of which is located in the central molecular zone (CMZ).  The proximity of NGC\,253 makes it possible to study the nuclear starburst at high resolution and sensitivity and it provides an important testbed for understanding the relationship between star formation in normal spirals, nuclear starbursts and merger-driven starburst systems \citep{Leroy+15}.  Observations of the OH and water maser transitions towards NGC\,253 have detected maser emission from each \citep{Turner+85,Henkel+04}, however, the class~I methanol emission is offset from the location of water masers (Fig.~\ref{fig:ngc253}).

\section{Maser-starburst connection}

The initial detection of the 36~GHz methanol emission in NGC\,253 has recently been confirmed in follow-up observations made with both the
very large array (VLA) \citep{Gorski+17} and the Australia Telescope Compact Array (ATCA) \citep{Ellingsen+17b}.  To date extragalactic class~I
methanol maser emission has been observed at high angular resolution towards three sources NGC\,253, NGC\,4945 \citep{McCarthy+17} and Arp\,220 \citep{Chen+15}; with tentative detections with single dish telescopes made towards a further 10 or so sources \citep{Wang+14,Chen+16}.  For the three sources where there is high resolution information on the distribution of the methanol masers, they are observed significantly offset from the nucleus of the galaxy (100s-1000s of pc), so the natural question to ask is - what do the class~I extragalactic masers trace?

It is widely thought that the bar-potential plays a critical role in driving the nuclear starburst in NGC\,253 \citep[e.g.][]{Garcia-Burillo+00} and interestingly, the south-western 36~GHz methanol maser emission lies at the point where the inner edge of the bar meets the central molecular zone (see Fig.~\ref{fig:ngc253}).  Sensitive, high-resolution observations of the molecular gas in NGC\,253 show that this is a region of wide spread low-velocity shocks \citep{Meier+15}.  The 36~GHz class~I transition is observed to be widespread in the CMZ of the Milky Way \citep{Yusef-Zadeh+13} and  Galactic class~I methanol maser emission is known to be associated with molecular outflows \citep[e.g.][]{Voronkov+06,Cyganowski+09}, expanding H{\sc ii} regions \citep[e.g.][]{Voronkov+10a}, cloud-cloud collisions \citep[e.g.][]{Sobolev92} and molecular cloud-SNR interactions \citep[e.g.][]{Pihlstrom+14}.  The two common factors which are present in all of these environments is the presence of cool molecular gas and low-velocity shocks.

Cool molecular gas is the basic fuel for star formation, while low-velocity shocks are the trigger for the star formation process.  Hence, early- and intermediate-stage starburst galaxies must have both of these present on scales which are much larger than that of typical Galactic star formation regions or the CMZ of the Milky Way.  However, the class~I methanol maser emission in NGC\,253 does not appear to be simply a scaled-up version of that typically observed in Galactic star formation regions.  The ATCA observations show that rather than being a blend of a large number of compact emission regions, the majority of the 36~GHz methanol emission in NGC\,253 is relatively diffuse and resolved out on spatial scales smaller than $\sim$50~pc \citep{Ellingsen+17b}.  This is similar to what is observed for OH and formaldehyde megamaser emission \citep{Baan+17} and in contrast to the very compact emission seen in water megamasers (see the paper by Baan in these proceedings).  Furthermore, the 44~GHz class I methanol maser emission in NGC\,253 is two orders of magnitude weaker than the 36~GHz masers \citep{Ellingsen+17b}, which is very different from what is observed in typical Galactic star formation regions \citep{Voronkov+14}.  When comparing the thermal molecular gas in NGC\,253 with the 36~GHz methanol maser emission, \citet{Ellingsen+17} observed the HC$_3$N (J=4-3) emission to be very similar in distribution, with some of the emission detected on 0.1$^{\prime\prime}$ scales.  These observations show that some of the HC$_3$N emission is masing.  This is the first detection of maser emission in this transition in any source, Galactic or extragalactic, and further evidence that the class~I methanol maser emission in NGC\,253 is a different phenomenon from that observed in Galactic sources.    

\section{Conclusions}

The 36~GHz class~I methanol masers appear to be an exciting new tool for investigations of molecular gas in extragalactic sources.  In NGC\,253 the emission appears to trace regions where cool molecular gas is undergoing low-velocity shocks, such as at the interface between the galactic bar and the central molecular zone.  Further investigation of the masing and molecular gas in other sources is required to determine their similarities and differences and hence obtain a more comprehensive understanding of the critical aspects which determine where strong extragalactic class~I methanol masers are likely to be found.  If the luminosity of the class~I methanol maser emission scales with the star formation rate \citep[as has been suggested by][]{Chen+16}, then it may be possible to detect much more distant sources in the 36~GHz transition and use this to investigate starburst galaxies.


\section*{Acknowledgements}
\noindent
I would like to thank the following collaborators who have made contributions to the work presented here, Xi Chen (Guangzhou University \& Shanghai Astronomical Observatory), Shari Breen (University of Sydney), Hai-Hua Qiao (Shanghai Astronomical Observatory), Willem Baan (ASTRON), Tiege McCarthy (University of Tasmania) and Maxim Voronokov (CSIRO Astronomy and Space Science).

\bibliography{/Users/sellings/tex/references.bib}



\end{document}